\documentclass[conference]{IEEEtran}
\IEEEoverridecommandlockouts
\usepackage{cite}
\usepackage{amsmath,amssymb,amsfonts}
\usepackage{algorithmic}
\usepackage[ruled,linesnumbered]{algorithm2e}
\usepackage{graphicx}
\usepackage{textcomp}
\usepackage{xcolor}
\usepackage{bm}
\usepackage{epstopdf}
\usepackage{balance}
\def\BibTeX{{\rm B\kern-.05em{\sc i\kern-.025em b}\kern-.08em
    T\kern-.1667em\lower.7ex\hbox{E}\kern-.125emX}}
\begin{document}

\title{On the Age of Information for AMP based Grant-Free Random Access\\
}

\author
{
    \IEEEauthorblockN{Dongliang Zhang}
    \IEEEauthorblockA{\textit{School of Electronics and} \\
    \textit{Information Technology}\\
    \textit{Sun Yat-Sen University}\\
    Guangzhou, China \\
    zhdliang@mail2.sysu.edu.cn}
    \and
    \IEEEauthorblockN{Jie Gong}
    \IEEEauthorblockA{\textit{School of Computer Science and} \\
    \textit{ Engineering}\\
    \textit{Sun Yat-Sen University}\\
    Guangzhou, China \\
    gongj26@mail.sysu.edu.cn}
    \and
    \IEEEauthorblockN{Xiang Chen}
    \IEEEauthorblockA{\textit{School of Electronics and} \\
    \textit{Information Technology}\\
    \textit{Sun Yat-Sen University}\\
    Guangzhou, China \\
    chenxiang@mail.sysu.edu.cn}
}

\maketitle

\begin{abstract}
    With the rapid development of Internet of Things (IoT), massive devices are deployed,
    which poses severe challenges on access networks due to limited communication resources.
    When massive users contend for access, the information freshness gets worse caused 
    by increasing collisions. It could be fatal for information freshness sensing scenarios,
    such as remote monitoring systems or self-driving systems, 
    in which information freshness plays a critical part.
    In this paper, by taking the Age of Information (AoI) as the primary performance indicator, 
    the information freshness using AMP-based grant-free scheme is investigated and compared with grant-based scheme. 
    Base on the analysis, a user scheduling strategy with sleep threshold
    and forcing active threshold is proposed to further reduce average AoI (AAoI). 
    Numerical results reveal that the AMP-based grant-free scheme can provide sufficient access capability with less pilot resources,
    and it is robust to the fluctuation of the number of active users. 
    That ensures that the AMP-based grant-free scheme can keep the AAoI at a low level. 
    It is also shown that the proposed threshold strategy can effectively improve the information freshness.
\end{abstract}
\begin{IEEEkeywords}
    Age of Information, Approximate Message Passing, Grant-free random access schemes, Threshold Strategy
\end{IEEEkeywords}
\section{Introduction}
With the boom of Internet of Things (IoT) technology, IoT devices are widely applied in daily lives and provide ubiquitous connectivity 
for machines or objects which may not require human interactions. Taking the convenience of the IoT technology, our lives are significantly
impacted and changing to a smarter form. 
The massive number of devices and the complexity of interconnection bring convenience to 
people's lives but also challenges to the communication systems with massive access requests. 
One of the key application scenarios of IoT is real-time monitoring and control, 
such as remote monitoring systems or self-driving systems, there is a high demand for the freshness of information.
It is challenging to enhance the radio access capacity to guarantee information freshness.

The International Telecommunication Union (ITU) has defined massive machine-type communications (mMTC) \cite{m1}. 
It has the characteristics of massive number of devices, a small amount of data transmission, low power consumption, 
and sporadic device active pattern\cite{m2}, i.e., at any specific time only 
a small fraction of devices are active. It means that devices are usually in a sleep state to save energy, 
and are activated only when triggered by external events. As a result, the base station (BS) needs to dynamically
identify the active users and receive the transmitted data. 

A common random access approach is the grant-based scheme which requires each device to compete for the 
limited orthogonal pilot resources. But in the massive IoT scenarios, many users cannot access the network due to the frequent
collisions caused by limited pilot resources and massive access requests. In these scenarios, the grant-free random access schemes 
will be a promising access approach. It allows devices to transmit their metadata and data to the BS without waiting for permission.

In the past decades, several grant-free schemes have been introduced and their performances are analyzed.
To serve the sporadically active users, irregular repetition slotted ALOHA (IRSA) was proposed 
which is a low-complexity grant-free random access \cite{m3},\cite{m4}.
In \cite{m5}, a low-complexity dynamic compressed sensing based multi-user sensing algorithm was exploited, 
by using the time correlation of the active users to estimate the current transmitted data.
In \cite{m6},\cite{m7}, the approximate message passing (AMP) algorithm is used to perceive the user's activity, and the BS performs channel estimation based on the received meta-information, 
and then decodes data with the estimated channel states.
Previous works focused on the throughput and the detection error probability, but the optimization of information freshness was not considered.

For real-time IoT applications, Age of Information (AoI) was introduced in \cite{m11}
to measure the freshness of data and used to enhance the real-time performance.
In \cite{m8}, a threshold method is introduced to improve the slotted ALOHA algorithm and gives calculation of AoI 
carried 
~\\ 
out by the algorithm.
In \cite{m9}, a novel early packet recovery method 
for IRSA is proposed which further reduces the average AoI (AAoI).
The AoI of the method was characterized and the problem of minimizing AAoI by optimizing over the normalized channel traffic and repetition distribution was considered.
However, the AoI analysis of the AMP-based grant-free random access method, which is a more efficient access method, has not been considered, 

This paper takes AoI as indicator to compare the efficiency between the grant-based schemes and the grant-free
schemes and characterizes the AAoI of the network using Markov chain.
Based on the AMP grant-free schemes, 
a user scheduling strategy with sleep threshold and forcing active threshold is proposed to promote the AAoI of the system 
and an algorithm to calculate the AAoI is given. The numerical results demonstrate the effectiveness of grant-free schemes with proposed strategy.

\section{System Model}
Consider the uplink of a single-cell cellular network with BS equipped with a single antenna and connected to $N$ potential users
which denotes by the set $\mathcal{N} = \{ 1, ..., N\}$. In each time slot, each user enters the active state with probability $\epsilon$.
Each active user accomplishes its data transmission within a time slot, and users are synchronized. 
Define the user active state function for $\forall n \in  \mathcal{N} $ as
\begin{equation}
	a_n=
	\begin{cases}
		1,& \text{if user $n$ is active} \\
		0,& \text{otherwise}
	\end{cases}
\end{equation}
Then, $a_n$ follows Bernoulli distribution so that $P(a_n=1)=\epsilon$ and $P(a_n=0)=1-\epsilon, \forall n$.
And the average number of active users in each time slot is $K = N\epsilon$.
The channel from user $n$ to the BS is $h_n = \sqrt{\beta _n}g_n$, where $\beta _n$ denotes the Rayleigh fading component, 
and $g_n$ denotes the path-loss component. 
Block fading channel is assumed and $h_n$ remains unchanged in each time slot.

\subsection{Grant-based Model}
The main performance degradation of random access lies in the collision caused by contention for limited orthogonal sequences
to get aligned with the BS\cite{m10}. It is assumed that if there is no contention, the transmission will be successful.
The grant-based scheme applied in this paper is slotted ALOHA. Define the length of pilot sequence is L, so the number of orthogonal sequences is also L.
In each time slot, active users randomly request for the $L$ orthogonal sequences. When an orthogonal sequence is selected by only one user, the random access is successful.
However, if two or more users select the same sequence at the same time, a conflict occurs and random access fails.
Assume the length of transmitted data is $D$.
It takes one time slot to transmit both pilot and data.

\subsection{Grant-free Model}
As the user's data transmission is sporadic, it can be modeled as a compressed sensing problem.
The BS initially assigns one dedicated pilot for each user 
$\bm{a}_n \in \mathbb{C}^{L\times 1} \sim \bm{CN}(0,1/L)$ with $\|\bm{a}_n\|^2 = 1$, where L denotes the length of the pilot.
The perception matrix is formed as $\bm{A}=[\bm{a}_1,...,\bm{a}_N]$.
The received signal at the BS is then
\begin{equation}
	\bm{y} = \sqrt{\xi}\bm{Ax} + \bm{z},
\end{equation}
where y is the received signal, $\xi$ is the total energy of all active users' pilots, 
$\bm{x}=[x_1,...,x_N]^T$ with $x_n=a_n h_n$, denotes a set of users' signal,
and $\bm{z} \in \mathbb{C}^{L\times 1} \sim \bm{CN}(0,1)$ is the independent additive white Gaussian noise.
The task for the BS is to use the received signal and the known perception matrix to detect the active users, 
which is challenging and most resource-consuming part. Thus the subsequent decoding is assumed error-free for fair comparison.
This paper adopts the AMP algorithm with MMSE denoiser to implement the detection and the AMP algorithm 
proceeds at each iteration as \cite{m6},\cite{m7}:
\begin{equation}
    \bm{x}_{n}^{t+1}=\bm{\eta}_{t,n}((\bm{r}^{t})^{\bm{H}}\bm{a}_{n}+x_{n}^{t}),
    \label{xhat}
\end{equation}
\begin{equation}
    \bm{r}^{t+1}=\bm{y}-\bm{Ax}^{t+1}+\frac{N}{L}\bm{r}^{t}\sum_{n=1}^N\frac{\bm{\eta}_{t,n}^{'}((\bm{r}^{t})^{\bm{H}}\bm{a}_{n}+x_{n}^{t})}{N},
    \label{residual}
\end{equation}
where $t=0,1,...$ is the index of the iteration, $L$ is the length of the pilot sequence, $\bm{\eta}_{t,n} ():\mathbb{C} \to \mathbb{C}$ is the denoiser,
$\bm{\eta}_{t,n}^{'}$ is the first-order derivative of the denoiser function.

\subsection{Age of Information}
AoI is proposed to quantify the freshness of received information in the network, 
which is defined as the time elapsed since the generation of the last successfully received message containing update information.
It uses the perspective of examining the freshness of information to promote system performance, which is important for 
real-time communications or control systems.
Define $t_i$ as the time when the last successfully received message was generated, and the AoI $\bm{\Delta}[t]$ is 
\begin{equation}
    \bm{\Delta}[t] = t - t_i.
\end{equation}
It indicates that when the information is updated, the AoI will be updated to $t_i ^{'} - t_i$, 
where $t_i ^{'}$ is the update time stamp, and then $\bm{\Delta}[t]$ will increase linearly over time until the next update.
To analyze the performance of the entire network over a long period of time, it is necessary to calculate the AAoI of the network:
\begin{equation}
    \bar{A} =  \sum^N_{i=1} \omega_i \bar{A}_i[t],
\end{equation}
where $\omega_i$ is the weight of each user's AoI.

\section{Analysis and Design of AMP-Based Grant-free scheme}
In this section, the AAoI of the system is characterized by using Markov chain analysis based on the user access success rate.
Based on the original AMP-based access method, a user scheduling strategy with sleep threshold and forced active threshold is proposed.
And the user active probability and the AAoI of the system is analyzed.

\subsection{AAoI Analysis Based on Access Success Rate}
Assume independently and identically distributed active probability and access success rate for each user.
Hence, the AAoI of the entire system is equal to the AAoI of a single user.

As user's AoI state only depends on the state in the previous time slot, it can be modeled as a Markov chain
and the state set is
\begin{equation}
    A[t]=
    \begin{cases}
        1,& \text{update successfully} \\
        A[t-1]+1,& \text{otherwise}
    \end{cases}
\end{equation}

The transition probability matrix is 
\begin{equation}
    P =
    \begin{pmatrix}
        p_u & p_e & & & \\
        p_u & & p_e & & \\
        \vdots & & &\ddots & 
    \end{pmatrix},
\end{equation}
where $p_u = \epsilon \rho$ is the access success rate and $p_e = 1-p_u$.
$\epsilon = K/N$ is the user active probability and $\rho$ is the access success rate. 
When update is success, $A[t]$ is reset to 1.
By solving the equilibrium equation $\bm{\hat{\pi}} = \bm{\hat{\pi}} P$, where $\bm{\hat{\pi}} = (\hat{\pi}_1, \hat{\pi}_2, ...)$, the steady state probability is 
\begin{equation}
	\hat{\pi}_n = 
    \begin{cases}
        p_u,& n=1\\
        p_u p_e^{n-1},& n>1
    \end{cases}
\end{equation}
where the index of each state represents the value of its current AoI.
Then the AAoI is:
\begin{align}
    \bar{A} = \sum_{n=1}^{\infty}n\hat{\pi}_n = p_u\sum_{n=0}^{\infty}p_e^{n}(n+1) = \frac{1}{p_u},
\end{align}
It means that AAoI of a single user is the reciprocal of the update success rate when the system enters steady state.
Considering the symmetry between users, the AAoI of the system is also the reciprocal of the update success rate.
In simulation experiments, this equation can be used to predict the AAoI of a system with limited runtime.

\subsection{User Scheduling Strategy with Sleep Threshold and Forcing Active Threshold}
In order to reduce the AAoI of the network, consider how to provide more effective update information 
by adjusting the user scheduling under the same constraints.
The intuitive way is to request users who have not updated for a long time to update immediately,
but this will increase the probability of user activity and increase the overall transmission power consumption of the network.
In order to keep the user's active probability unchanged, the sleep threshold strategy is added, 
and the user's active probability can be kept unchanged through the coordination of the two thresholds.
By avoiding the users who have updated recently to update repeatedly, 
and letting the users who have not updated for a long time to update information,
it is expected to reduce the AAoI of the system.

Denote the sleep threshold by $\underline{\theta}$, the forced active threshold by $\overline{\theta}$, 
the user's active probability in the unforced state by $\epsilon$
and the access success rate of the BS by $\rho$.
Taking the time interval from user's last activity to now $T[t]$ as its state. 
When the time interval is less than or equal to the sleep threshold, 
the user's activity probability is set to 0. 
When the time interval is greater than or equal to the forced active threshold, 
the user will be forced to be active, the update information will be sent to the BS immediately.
But in the unforced state, the user will randomly update the information based on the preset active probability. 
Since the user performs synchronization with the BS, 
it is aware of the active interval state of itself, and use its own state to make decision.

Due to users' symmetry, the analysis result of the activity probability of single user can be applied to all users.
Because the time interval of the next slot is only related to the interval in the current slot, it can be modeled as Marko Chain 
and user's state is 
\begin{equation}
	T[t]=
	\begin{cases}
		1,\quad \quad \quad \quad \quad \underline{\theta} < T[t-1] < \overline{\theta}\text{, user is active} \\
        \quad \text{Or } T[t-1] = \overline{\theta}\text{, user is forced to be active}.\\
        T[t-1]+1, \ \underline{\theta} < T[t-1] < \overline{\theta}\text{, user is not active} \\
        \quad \text{Or } T[t-1] \leqslant  \underline{\theta} \text{, user is forced to sleep}
	\end{cases}
\end{equation}

Because when $T[t]\geqslant\overline{\theta}$, the user sends out data immediately and $T[t]$ is reset to 1.
So that the states $T[t]>\overline{\theta}$ is transient which can be omitted.
The state transition matrix is
\begin{equation}
    Q = 
	\begin{pmatrix}
		0 & 1 & & & & & \\
        \vdots & & \ddots & & & & \\
        0 & & & 1 & & & \\
        \epsilon & & & & 1-\epsilon & & \\
        \vdots & & & & & \ddots  & \\
        \epsilon & & & & & & 1-\epsilon \\
        1 & & &  & & & 0
	\end{pmatrix},
\end{equation}

By solving the equilibrium equation, the steady state probability is
\begin{equation}
    \label{pi1}
		\tilde{\pi} _n = 
		\begin{cases}
			\frac{1}{1+\underline{\theta}+\sum_{i=1}^{\overline{\theta}-\underline{\theta}-1}(1-\epsilon)^i},& n = 1\\
			\tilde{\pi}_{n-1},& 1 < n \leq \underline{\theta}+1\\
			(1-\epsilon)\tilde{\pi}_{n-1},& \underline{\theta}+1 < n \leq \overline{\theta}
		\end{cases}
\end{equation}
Since when a user becomes active in any time slot its next state will back to $T[t]=1$,
so the total active probability of the user $\epsilon$ under long-term operation is equal to the steady state probability $\pi1$, 
i.e., $\epsilon = \tilde{\pi}_1$.
To keep the total active probability of users unchanged, just keep $\tilde{\pi}_1$ unchanged with certain threshold pair.

Subsequently, the AAoI of the system can be obtained using the steady state probability of $T[t]$.
Even if the user enters the active state and sends information to the BS,
there is still a probability of $1-\rho$ that the BS cannot receive the updated information.
Therefore, it is difficult to obtain the state transition matrix of the AoI directly.
Consider using the steady state probability of the time interval state of single user obtained above to indirectly obtain the stability of the AoI state.
Let $i$ be the time interval of the user at the current time $t$, $i=T[t]$, 
and let $p_i$ be the active probability of the user in the current state:
\begin{equation}
    p_i = 
    \begin{cases}
        0,& i \leq \underline{\theta}\\
        \epsilon,& \underline{\theta} < i < \overline{\theta}\\
        1,& i = \overline{\theta}
    \end{cases}
\end{equation}	
For a single user in state $i$, let its AoI at the BS be $j$, 
$S[t]$ is the state of the user, where $S[t] = (A[t], T[t]) = (j, i)$.
Denote the steady state probability of $S[t] = (j, i)$ by $\pi_{j,i}$.
For user in $S[t] = (j,i)$,  there are three transfer possibilities, 
one is that the user is active and successfully updated, 
the other is that the user is active but the update fails, and the third is that the user is not active:
\begin{equation}
    P_r(S[t+1] | S[t]) = 
	\begin{cases}
		p_i \rho, &\text{if } S[t+1]=(1,1)\\
		p_i(1-\rho), &\text{if } S[t+1]=(j+1,1)\\
		1-p_i, &\text{if } S[t+1]=(j+1,i+1)
	\end{cases}
\end{equation}
When the user gets active and access the BS successfully, the next state must be $S[t+1] = (1,1)$.
Therefore, $\pi_{1,1}$ is 
\begin{equation}
	\pi_{1,1} = \sum_{i=1}^{\overline{\theta}}\tilde{\pi}_i p_i \rho.
\end{equation}
Since the states $S[t] = (1, i)$, $i > 1$ are transient, their steady-state probabilities are zero, i.e., $\pi_{1,i}=0$, $i>1$.
Therefore, the steady-state probability at $S[t] = (1, i)$, $i > 1$ is obtained,
and the steady-state probability at other AoI state $j$ can be obtained by the following function:
\begin{equation}
	\pi_{j,i} =
	\begin{cases}
		\sum_{n=\underline{\theta}+1}^{\overline{\theta}} \ p_n(1-\rho)\pi_{j-1,n}, i = 1 , j > 1\\
		(1-p_{i-1})\pi_{j-1,i-1}, 1<i\leq \overline{\theta} , j > 1
	\end{cases}
\end{equation}
We set a limit $\bar{T}$ for $j$, where $\bar{T}$ is the time when the system enters steady state, for simulation.
Let $\pi_j^{'}$ be the steady-state probability when the AoI state is $j$:
\begin{equation}
	\pi_j^{'} = \sum_{i=1}^{\overline{\theta}}\pi_{j,i}.
\end{equation}
The AAoI of single user over a long runtime can be obtained:
\begin{equation}
	\bar{A}^{'} = \sum_{j=1}^{\infty}j\pi_j^{'}.
\end{equation}
Because of the symmetry of users in the network, 
the AAoI of the system is equal to the AAoI of a single user.
The calculation process of $\bar{A}'$ is summarized as Algorithm 1.

\begin{algorithm}
    \caption{Calculation of the AAoI $\bar{A}'$}
    \label{algorithm1}
    \KwData{user active probability $p_i$, access success rate $\rho$, user steady state $\tilde{\pi}_i$}
    \KwResult{AAoI $\bar{A}'$}

    $\pi_{j,i} \leftarrow 0, $ for $\forall j, i$\;
    $\pi_{1,1} \leftarrow \sum_{i=1}^{\overline{\theta}}\tilde{\pi}_i p_i \rho$\;

    \For{$j \leftarrow 2$ \KwTo $\bar{T}$}
    {    
        \For{$i \leftarrow 1$ \KwTo $\bar{\theta}$}
        {
            
            \eIf{$i = 1$}
            {
                $\pi_{j,i} \leftarrow \sum_{n=\underline{\theta}+1}^{\overline{\theta}} \ p_n(1-\rho)\pi_{j-1,n}$\;

            }
            {
                $\pi_{j,i} \leftarrow (1-p_{i-1})\pi_{j-1,i-1}$\;
            }
        }
    }
    \For{$j \leftarrow 1$ \KwTo $\bar{T}$}
    {   
        $\pi_j^{'} \leftarrow \sum_{i=1}^{\overline{\theta}}\pi_{j,i}$\;
    }
    $\bar{A}^{'} \leftarrow \sum_{j=1}^{\bar{T}}j\pi_j^{'}$\;
\end{algorithm}

\section{Numerical Results}
In this section, AAoI comparison between grant-based method and AMP-based method is given and 
numerical examples of threshold strategy is shown.

\subsection{AAoI Comparison with Grant-based Scheme}

The default setup of the experiment is as follows. There are $N=2000$ users in the cell, and at any time slot, each user's active
probability is $\epsilon = 0.05$, and the length of user pilot is $L=200$. In each experiment, only a single variable is changed 
in order to find out the factors that have impact on the system AAoI.

\begin{figure}[htbp]
    \centering
    \includegraphics[width=0.49\textwidth]{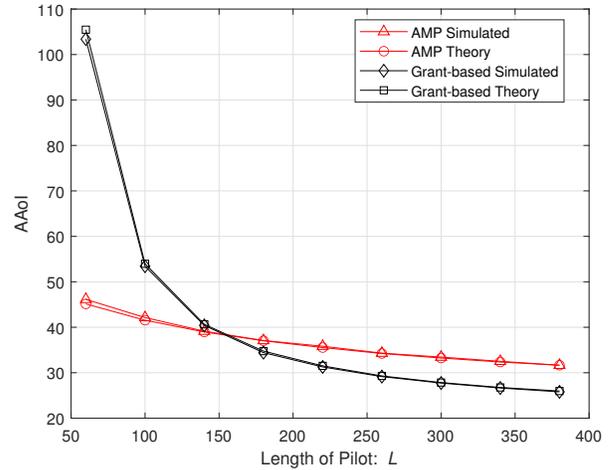}
    \caption{AAoI of systems with different length of pilot when $N = 2000$ and $\epsilon = 0.05$}
    \label{LDif}
\end{figure}

First we examine the effect of length of pilot and user's active probability. 
With a fixed $N$ and $\epsilon$, only change the pilot length from 40 to 380. Because active probability $\epsilon$ is fixed, 
AAoI only depends on the access success rate. 
When the length of the pilot sequence increases, it brings additional orthogonal sequences 
significantly promoting the access success rate of the grant-based method, resulted in a significant decrease in AAoI, 
as shown in Fig. \ref{LDif}.
\begin{figure}[htbp]
    \centering
    \includegraphics[width=0.49\textwidth]{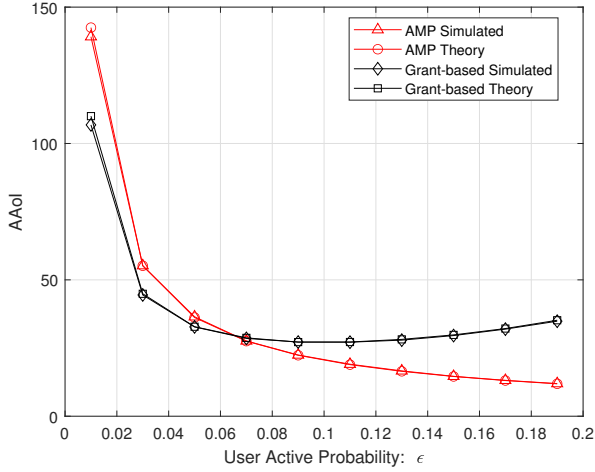}
    \caption{AAoI of systems with different user active probability when $N = 2000$ and $L = 200$}
    \label{eDif}
\end{figure}
For the AMP-based grant-free method, the AAoI decreasing slowly because its access success rate is hardly affected.

While with a fixed $N$ and $L$, but a changing active probability $\epsilon$ from 0.02 to 0.20, things get a little different.
Notice that the update information only comes from the active users, so the user active probability denotes 
the rate of update information generation. When user active probability approaching zero, the AAoI approaches 
infinity with both access methods.
When the user active probability increases, sufficient update information first keeps AAoI going down 
but deteriorates the access success rate due to the limited length of pilot.
For grant-based method, as access success rate drops quickly, Fig. \ref{eDif} shows a trend of increasing first and then decreasing.
But for AMP-based method, as access success rate drops slowly, its AAoI keeps falling. 
The above two examples reveal that the increasing collision greatly reduce AAoI of grant-based method under limited pilot.
But the AMP-based method can use shorter pilot for better AAoI.
Because it uses the covariance matrix obtained in the iteration to predict the error, which is robust to $\epsilon$ and $L$.

\begin{figure}[htbp]
    \centering
    \includegraphics[width=0.49\textwidth]{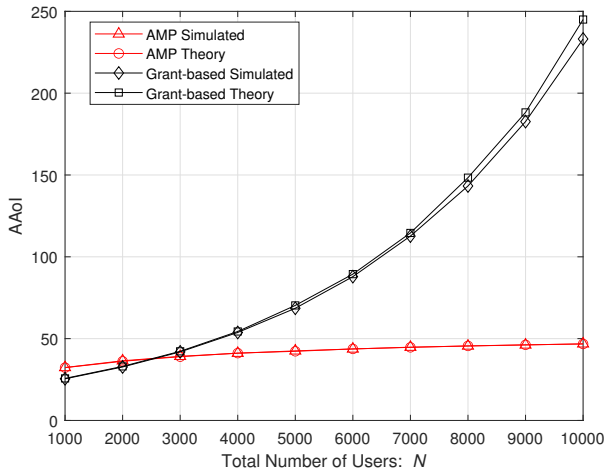}
    \caption{AAoI of systems with different total number of users $N$ when $\epsilon = 0.05$ and $L = 200$}
    \label{NDif}
\end{figure}

Second we examine the effect of the total number of users $N$.
The result is shown in Fig. \ref{NDif}.
With fixed $L$ and $\epsilon$,only change the total number of users from 1000 to 10000.
As the increasing of user's number leads to more active users, 
increasing collisions makes access success rate of grant-based method decrease.
But AMP-based grant-free method can still maintain slower AAoI,
because it makes full use of the increasing update information.
It shows that the AMP-based grant-free method is robust in the limited pilot massive access scenario.
\begin{figure}[htbp]
    \centering
    \includegraphics[width=0.49\textwidth]{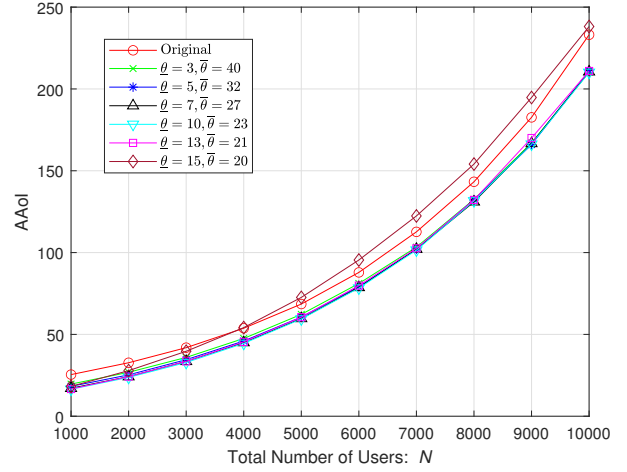}
    \caption{AAoI of systems with threshold strategy grant-based method when $\epsilon = 0.05$ and $L = 200$}
    \label{grantbased strategy}
\end{figure}
\begin{figure}[htbp]
    \centering
    \includegraphics[width=0.5\textwidth]{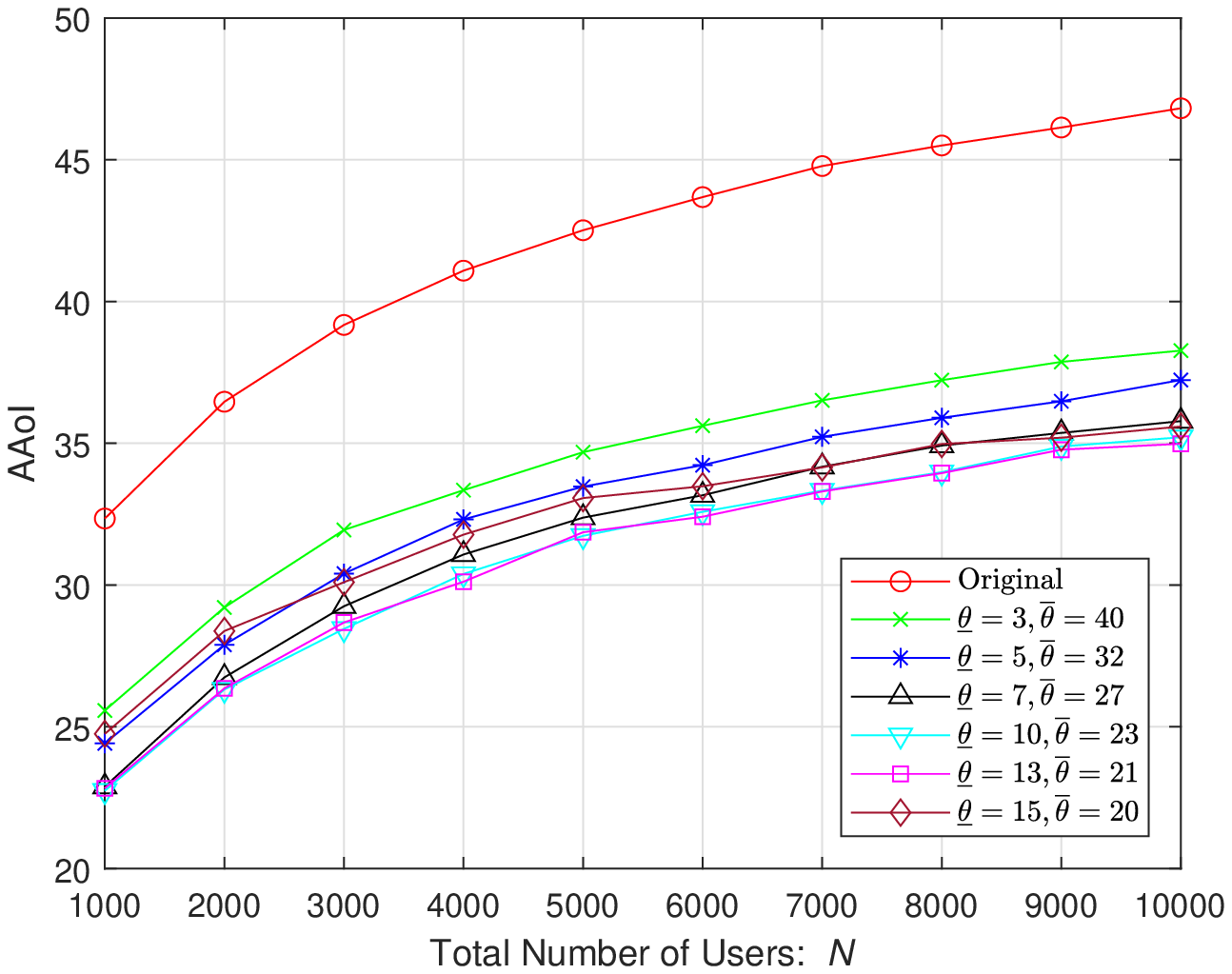}
    \caption{AAoI of systems with threshold strategy AMP-based grant-free method when $\epsilon = 0.05$ and $L = 200$}
    \label{AMP strategy}
\end{figure}
\begin{figure}[htbp]
    \centering
    \includegraphics[width=0.5\textwidth]{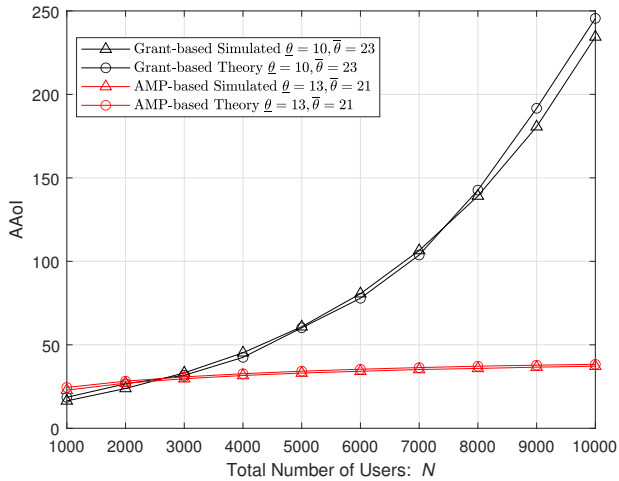}
    \caption{Comparison of the best case of the two access method }
    \label{best}
\end{figure}
\subsection{Threshold Strategy Performance}
With fixed $L=200$ and $\epsilon=0.05$, only change the total number of users from 1000 to 10000.
With $\pi _1$, the threshold pairs can be obtained when the user activity probability is kept at 0.05.
Extract several pairs for simulation experiments.
The results of grant-based method and results of AMP-based grant-free method are shown in Fig. \ref{grantbased strategy} and Fig. \ref{AMP strategy} respectively.
The AAoI of both methods with threshold strategy decrease and the AAoI of AMP-based method reduces significantly.
In the optimal case, there is an average 23\% reduction for the AMP-based method.
The main role is the forced active threshold, which reduces the accumulation of outdated users in the network.
The sleep threshold is used to maintain the user active probability without losing too much information freshness.
In Fig. \ref{best},  we compare the best cases of the two access methods.
It shows that the AMP-based method can provide higher information freshness with threshold strategy.

\section{Conclusion}
This paper shows that the AMP-based grant-free method can better maintain the information freshness in massive user access scenario
under the constraint of short pilot numbers. 
And a user scheduling strategy with sleep threshold and forcing active threshold 
is proposed on the basis of AMP-based method and has been verified that it can effectively reduce the AAoI of the original system.

\balance

\end{document}